\documentclass{article}

\usepackage[english]{babel}
\usepackage[a4paper,top=2cm,bottom=2cm,left=3cm,right=3cm,marginparwidth=1.75cm]{geometry}

\usepackage{amsmath}
\usepackage{algorithm}
\usepackage{algpseudocode}
\usepackage{appendix}
\usepackage{epigraph}
\usepackage[scaled]{helvet}

\usepackage{graphicx}
\usepackage[colorlinks=true, allcolors=blue]{hyperref}
\usepackage{blindtext}
\usepackage{csquotes}
\usepackage[backend=biber,style=numeric]{biblatex}
\usepackage{multirow}
\usepackage{authblk}
\usepackage[all]{nowidow}

\begin{document}
\title{Interleaved Snowballing: Reducing the Workload of Literature Curators}
\author{Ralf Stephan}
\affil{Institute for Globally Distributed Open Research and Education (IGDORE)

\vspace{0.3cm}
\footnotesize Address correspondence to: Ralf Stephan, Institute for Globally Distributed Open Research and Education (IGDORE), https://igdore.org. E-mail: ralf.stephan@igdore.org
}
\maketitle

\begin{abstract}
We formally define the literature (reference) snowballing method and present a refined version of it. We show that the improved algorithm can substantially reduce curator work, even before application of text classification, by reducing the number of candidates to classify. We also present a desktop application named LitBall\cite{litball} that implements this and other literature collection methods, through access to the Semantic Scholar academic graph\cite{Kinney2023TheSS}.
\end {abstract}

\begin{minipage}[t][3\baselineskip][t]{\linewidth}
\epigraph{\textsc{to interleave:} to put layers or flat pieces of something between layers or flat pieces of something else}{}
\end{minipage}
\section*{Introduction}
Snowballing is the method of choice to get a complete list of literature on a specific topic, and is applied by authors of systematic reviews, especially scoping reviews\cite{choong2014automatic,greenhalgh2005effectiveness,shemilt2014pinpointing,wohlin2014guidelines}.
Software that helps with snowballing is scarce and depends on the availability of a specific academic graph (AG) to work at all; as to recent implementations, we could only find Snowglobe\cite{mcweeny2021snowglobe} and Paperfetcher\cite{Pallath_2022}. The first uses the Microsoft Academic Graph (MAG)\cite{sinha2015overview}, which is no longer online, and the second uses Open Citations\cite{peroni2020opencitations} and CrossRef\cite{hendricks2020crossref}. We decided to write our own application, LitBall\cite{litball}, which uses the Semantic Scholar (S2) AG\cite{Kinney2023TheSS} that is arguably the largest open AG at the moment (Google Scholar AG lacks an API).

However, even with the snowballing process being automated, the number of articles that, in the end, have to be filtered out, can be enormous. If the filtering is done by hand/eye, it amounts to the bulk of the workload before the literature can be curated. Even using an ML classifier\cite{van2021symbals}, the classifier needs to be trained on the specific task to be useful, and it will never have a~100 percent hit rate. In this paper, we offer a refined snowballing algorithm that gives the same result as the original one, but reduces eyeballing by factors of up to ~$\approx$~6.

\section*{Definitions}
Literature curation requires the preliminary choice which scientific articles are \textit{curatable} at all, by looking at the title, the abstract (if available), or summaries of this information, like TLDR's from Semantic Scholar, or the full text. Later actual curation will use the
full text. Here. $F$ denotes the filtering process that decides which articles are curatable.
$$
\textbf{Curatable}(\textbf{article}) = 
F(\textbf{title}, \textbf{abstract}, \textbf{summary}, \textbf{full text}) =
\begin{cases}
    1  \\
    0
\end{cases}
$$
An \textbf{academic graph}(AG) is a directed acyclic graph, consisting of nodes (academic articles) and edges (references and citations).
Ideally, the set of all curatable articles $\mathcal{C}$ on a specific topic is determined over the complete academic graph $\mathbf{CAG}$ but, in reality, the graph used is a subset of $\mathbf{CAG}$, so $\mathcal{C}$ depends on which AG is used.
$$
\mathcal{C}(\text{AG},F) = \left\{\textbf{article}\in \text{AG}\;\middle|\;F(\textbf{article}) = 1\right\}
$$
Most of the time the subgraph C of AG, having nodes from $\mathcal{C}$ is connected but this
is not always the case.

We define the filter function $F$ as the conjunction of a simple pattern matching
function $F_1$ that takes a \textbf{regular or logical expression} given by the curator, and a second
function $F_2$ that effectively tries to filter out off-topic and non-curatable articles in the output
of $F_1$.
$$
F = F_2 \circ F_1 
$$
$F_2$ is performed by the curator or, increasingly, by text classification systems
benefiting from recent progress in machine learning. Still, specialized systems need training
data, and the accuracy is never~$100$\%, so the results have to be checked. This means the
amount of work by the curator performing~$F_2$ depends on and is proportional to the size
of~$\mathcal{M}_1=\{\textbf{article}\;|\;F_1(\textbf{article}=1\}$, i.e., what comes out of~$F_1$.
$$
\#\textbf{Work} \,\propto\, \#\mathcal{M}_1 = \,\#\left\{\textbf{article}\in \text{AG}\,\middle|\,F_1(\textbf{article}) = 1\right\}
$$
Using the \textbf{full text} of articles would improve the accuracy of classifiers but full text
in bulk is not available for the majority of articles. Still, improving classification 
accuracy for the subset of the AG which is available in bulk is an interesting option.

\section*{Snowballing and Interleaved Snowballing}
In this section we will assume that the set of articles suited for curation~$\mathcal{C}$ (the goal of the literature search) is connected. The idea of snowballing methods is then that it suffices to start with any node~$\in\mathcal{C}$ as part of~AG, and doing a breadth-first graph walk on~AG will soon encounter all members of~$\mathcal{C}$, by stopping the walk at nodes that are not~$\in\mathcal{C}$ (as determined by $F$), followed by backtracking\cite{choong2014automatic}.

If the AG is seen as having directed edges, the walk will only use references of nodes---this is called \textbf{backward snowballing}. But if the reference edges of~AG are seen as bidirectional, it will include the citations for any node (\textbf{forward snowballing}, e.g.\cite{felizardo2016using}). Only a combination of backward and forward snowballing can succeed to visit all members of~$\mathcal{C}$ in~AG.

Now, since the function $F$, that determines membership in~$\mathcal{C}$, consists of two consecutive filters, it matters when both filters are applied. Usually, literature snowballing applies pattern matching~($F_1$) during the snowballing rounds, and supervised classification~$(F_2)$ after the snowball is done:

\begin{algorithm}
\caption{Forward/Backward Snowballing}\label{alg:sn}
\begin{algorithmic}
\Require academic graph AG,\\ set of starting nodes $\mathcal{S}=\{s_1, s_2,\dots, s_i\}$,\\ pattern matching function~$F_1$,\\ supervised filter function~$F_2$
\State set of accepted nodes $\mathcal{A}_0 = \emptyset$\\ set of rejected nodes~$\mathcal{R}_0=\emptyset$
\State $\mathcal{A}_1 \gets \mathcal{S}$
\State $n \gets 1$
\While{$\mathcal{A}_n \supset \mathcal{A}_{n-1}$}
\State $\mathcal{A}_{n+1} \gets \mathcal{A}_n \cup \bigl\{\text{node} \in \text{AG}\,\bigm|\,\exists \,\text{edge}(\text{node},a) \,\wedge\, a\in\mathcal{A}_n \,\wedge\, F_1(\text{node})=1 \bigr\} $
\State $n \gets n+1$
\EndWhile
\State $\mathcal{M}_1 = \mathcal{A}_n$
\State $\mathcal{C} = \bigl\{a\in\mathcal{M}_1 \,\bigm|\, F_2(a) = 1 \bigr\}$
\end{algorithmic}
\end{algorithm}
We present an improved version of algorithm~\ref{alg:sn} that reduces the number of calls to~$F_2$ significantly, see~alg.~\ref{alg:isn}. The difference to algorithm~\ref{alg:sn} is that~$F_2$ is already applied during the expansion loop, reducing the increase of~$\mathcal{A}_{n+1}$. The number of calls to~$F_2$ is smaller than in algorithm~\ref{alg:sn} by a factor of (usually) up to $\approx$~6, see the examples in Table~\ref{tab:1}.

\begin{algorithm}
\caption{Interleaved Snowballing}\label{alg:isn}
\begin{algorithmic}
\Require academic graph AG,\\ set of starting nodes $\mathcal{S}=\{s_1, s_2,\dots, s_i\}$,\\ pattern matching function~$F_1$,\\ supervised filter function~$F_2$
\State set of accepted nodes $\mathcal{A}_0 = \emptyset$\\ set of rejected nodes~$\mathcal{R}_0=\emptyset$
\State $\mathcal{A}_1 \gets \mathcal{S}$
\State $n \gets 1$
\While{$\mathcal{A}_n \supset \mathcal{A}_{n-1}$}
\State $\mathcal{A}_{n+1} \gets \mathcal{A}_n \cup \bigl\{\text{node} \in \text{AG}\,\bigm|\,\exists \,\text{edge}(\text{node},a) \,\wedge\, a\in\mathcal{A}_n \,\wedge\, F_1(\text{node})=1 \,\wedge\, F_2(\text{node})=1\bigr\} $
\State $n \gets n+1$
\EndWhile
\State $\mathcal{C} = \mathcal{A}_n$
\end{algorithmic}
\end{algorithm}

Since the filter~$F_2$ is often applied by humans through supervision (e.g.~eyeballing of titles or abstracts) the algorithm could also be called \textit{Snowballing with Interleaved Supervision}.

\subsection*{Examples for reduction of $F_2$}
In order to get an overview of the improvement interleaved snowballing provides, we compare the number of applications of~$F_2$ by both algorithm~1 and~2 on the same search, where each search is defined by the academic graph (S2\cite{Kinney2023TheSS} in all cases), and identical parameters~$\mathcal{S}$, $F_1$, and~$F_2$ for both runs.

We used LitBall, which implements both algorithms: first using algorithm~2, in our daily work as biocurator, to search for lab-experimental articles and reviews. After that we simply ran algorithm~1 with the same parameters and recorded the number of "accepted" papers. The details are in Table~\ref{tab:1}.

A further result of interest was the number of accesses to the AG to get the paper data, which is, in both algorithms, $\#\mathcal{A}_n + \#\mathcal{R}_n$. The reduction factor for this number is of the same order of magnitude than that for~$F_2$. However, cost is not a problem if the AG API has a bulk access option.

\begin{table}[h]
    \centering
    \begin{tabular}{c|r|r|r|r|r|r}
        Name & Alg.1 $\#F_2$ & Alg.2 $\#F_2$ & Alg.1/Alg.2 & Alg.1 $\#\mathcal{A}_n + \#\mathcal{R}_n$ & Alg.2 $\#\mathcal{A}_n + \#\mathcal{R}_n$ & Alg.1/Alg.2 \\ \hline
        SLC25A10  & $95$ & $49$ & $\mathbf{1.9}$ & $5491$ & $1116$ & $\mathbf{4.9}$\\
        CSKMT & $258$ & $42$ & $\mathbf{6.1}$ & $11644$ & $513$ & $\mathbf{22.7}$ \\
        TAT & $>1000$ & $286$ & N/A & $>10000$ & $1273$ & N/A \\
        HGD & $265$ & $133$ & $\mathbf{2.0}$ & $658$ & $6859$ & $\mathbf{10.4}$ \\
        GSTZ1 & $113$ & $93$ & $\mathbf{1.2}$ & $5260$ & $622$ & $\mathbf{8.5}$ \\
        HPD & $689$ & $125$ & $\mathbf{5.5}$ & $22900$ & $490$ & $\mathbf{46.7}$ \\
        FAH & $427$ & $150$ & $\mathbf{2.8}$ & $18076$ & $559$ & $\mathbf{32.3}$ \\
        PDSS1,2 & $75$ & $49$ & $\mathbf{1.5}$ & $3915$ & $526$ & $\mathbf{7.4}$ \\
        COQ2 & $109$ & $53$ & $\mathbf{2.1}$ & $4000$ & $527$ & $\mathbf{7.6}$ \\
        COQ6 & $54$ & $44$ & $\mathbf{1.2}$ & $2509$ & $715$ & $\mathbf{3.5}$ \\
        COQ3 & $27$ & $21$ & $\mathbf{1.3}$ & $1710$ & $472$ & $\mathbf{3.6}$ \\
        COQ5 & $22$ & $15$ & $\mathbf{1.5}$ & $1046$ & $374$ & $\mathbf{2.8}$ \\
        COQ7 & $75$ & $44$ & $\mathbf{1.7}$ & $4018$ & $706$ & $\mathbf{5.7}$ \\
    \end{tabular}
    \caption{Results of snowballing processes in the S2 AG. Searches were for lab-experimental articles about specific proteins, starting with~1-5 articles referenced in UniProt. For exact regular expressions used see the appendix. In the case of TAT the search was stopped before completion, as the numbers in each cycle of algorithm~1 kept growing exponentially. This is a sign that the regex used for~$F_1$ was too broad. }
    \label{tab:1}
\end{table}

\section*{LitBall: the application}
LitBall\cite{litball} is a JVM desktop app for systematic literature collection using the Semantic Scholar AG. LitBall offers several search methods: expression search, snowballing, interleaved snowballing, and similarity search. It saves the state of snowball processing steps in a local database and all retrieved graph data. Output can be visualized or exported as a database for import in any spreadsheet.

LitBall uses Kotlin/Compose. There are binaries for Linux/Windows/Mac, made with the help of Conveyor. If you use LitBall, please give us feedback, especially on Win/Mac, as we only test on Linux.

\section*{Discussion}
We have formalized the snowballing algorithm and presented both an improvement and its implementation.
Through Semantic Scholar, the research community is lucky to have open and free access to an academic graph with metadata on $>200$ million documents and $>2.4$B citation edges (as of 2022\cite{Kinney2023TheSS}). The success of snowballing relies on the completeness of the underlying AG and, so, the S2 AG is a natural choice for applications using this method.

Early attempts on replacing the eyeballing part of snowballing with an ML classifier (e.g. SYMBALS \cite{van2021symbals}) were made, and we also tried to include a classifier interface with LitBall\cite{ralf_stephan_2023_8388963}. The work associated with curating the training data for this classifier, however, was substantial, with the topic covered being only a tiny part of possible topics that come up in biocuration, for example. The biggest hurdle for us is the general nonavailability of full text for articles, because only with full text lab-experimental articles can be reliably recognized. It may be possible to approximate the ideal by using large language models on abstracts, but then, abstracts aren't generally available, either.

Usage of snowballing is not confined to authors of systematic reviews. In our biocuration work for Reactome\cite{milacic2024reactome}, we daily use LitBall in interleaved snowballing, but also in other modes. For the frequent curator, it should be an important tool, but only one of many.

\printbibliography
\clearpage
\appendix
\section*{Appendix - Expressions used for filtering (case-insensitive, word-boundary)}\label{app}

\begin{table}[h]
    \centering
    \renewcommand{\arraystretch}{1.2}
    \begin{tabular}{c|p{5in}}
    CSKMT & CSKMT $\vee$ METTL12 $\vee$ CS-KMT $\vee$ Methyltransferase-like \\
    TAT & TAT $\vee$ tyrosine aminotransferase $\vee$ L-tyrosine:2-oxoglutarate \\
    HGD & HGD $\vee$ Homogentisate 1.2-dioxygenase $\vee$ Homogentisate dioxygenase $\vee$ HGO $\vee$ Homogentisate oxygenase $\vee$ Homogentisic acid oxidase $\vee$ Homogentisicase \\
    GSTZ1 & Maleylacetoacetate isomerase $\vee$ MAAI $\vee$ GSTZ1 $\vee$ GSTZ1-1 $\vee$ Glutathione S-transferase zeta 1 \\
    HPD & HPD $\vee$ 4-hydroxyphenylpyruvic acid dioxygenase $\vee$ 4-hydroxyphenylpyruvate dioxygenase $\vee$ 4HPPD $\vee$ HPPDase $\vee$ PPD \\
    FAH & FAH $\vee$ Fumarylacetoacetase $\vee$ Beta-diketonase $\vee$ Fumarylacetoacetate hydrolase \\
    PDSS1,2 & PDSS1 $\vee$ PDSS2 $\vee$ All trans-polyprenyl-diphosphate synthase $\vee$ Decaprenyl pyrophosphate synthase $\vee$ Decaprenyl-diphosphate synthase $\vee$ Solanesyl-diphosphate synthase $\vee$ Trans-prenyltransferase $\vee$ TPT $\vee$ TPT1 $\vee$ TPT2 $\vee$ DPS1 $\vee$ DPS2 $\vee$ TPRT $\vee$ C6orf210 \\
    COQ2 & COQ2 $\vee$ 4-hydroxybenzoate polyprenyltransferase $\vee$ 4-HB polyprenyltransferase $\vee$ Para-hydroxybenzoate--polyprenyltransferase $\vee$ PHB:PPT $\vee$ PHB:polyprenyltransferase $\vee$ CL640 \\
    COQ6 & Coenzyme Q10 monooxygenase 6 $\vee$ COQ6 $\vee$ CGI-10 \\
    COQ3 & COQ3 $\vee$ 3-demethylubiquinol 3-O-methyltransferase $\vee$ Polyprenyldihydroxybenzoate methyltransferase $\vee$ UG0215E05 \\
    COQ5 & COQ5 $\vee$ 2-methoxy-6-polyprenyl-1,4-benzoquinol methylase \\
    COQ7 & COQ7 $\vee$ 5-demethoxyubiquinone hydroxylase $\vee$ DMQ hydroxylase $\vee$ Timing protein clk-1 homolog \\
    \end{tabular}
\end{table}
\end{document}